\documentclass[lettersize,journal]{IEEEtran}
\usepackage{amsmath,amsfonts}
\usepackage{algorithmic}
\usepackage{algorithm}
\usepackage{array}
\usepackage[caption=false,font=normalsize,labelfont=sf,textfont=sf]{subfig}
\usepackage{textcomp}
\usepackage{stfloats}
\usepackage{url}
\usepackage{verbatim}
\usepackage{graphicx}
\usepackage{cite}

\begin{document}

\title{SST-ReversibleNet: Reversible-prior-based Spectral-Spatial Transformer for Efficient Hyperspectral Image Reconstruction}

\author{Zeyu Cai, Jian Yu, Ziyu Zhang, Chengqian Jin, Feipeng Da
}

\markboth{Journal of \LaTeX\ Class Files,~Vol.~14, No.~8, August~2021}%
{Shell \MakeLowercase{\textit{et al.}}: A Sample Article Using IEEEtran.cls for IEEE Journals}

\IEEEpubid{0000--0000/00\$00.00~\copyright~2021 IEEE}

\maketitle

\begin{abstract}
    Spectral image reconstruction is an important task in snapshot compressed imaging. This paper aims to propose a new end-to-end framework with iterative capabilities similar to a deep unfolding network to improve reconstruction accuracy, independent of optimization conditions, and to reduce the number of parameters. A novel framework called the reversible-prior-based method is proposed. Inspired by the reversibility of the optical path, the reversible-prior-based framework projects the reconstructions back into the measurement space, and then the residuals between the projected data and the real measurements are fed into the network for iteration. The reconstruction subnet in the network then learns the mapping of the residuals to the true values to improve reconstruction accuracy. In addition, a novel spectral-spatial transformer is proposed to account for the global correlation of spectral data in both spatial and spectral dimensions while balancing network depth and computational complexity, in response to the shortcomings of existing transformer-based denoising modules that ignore spatial texture features or learn local spatial features at the expense of global spatial features. Extensive experiments show that our SST-ReversibleNet significantly outperforms state-of-the-art methods on simulated and real HSI datasets, while requiring lower computational and storage costs. https://github.com/caizeyu1992/SST
\end{abstract}


\section{Introduction}
\label{sec:intro}

\begin{figure}[t]
  \centering
   \includegraphics[width=0.9\linewidth]{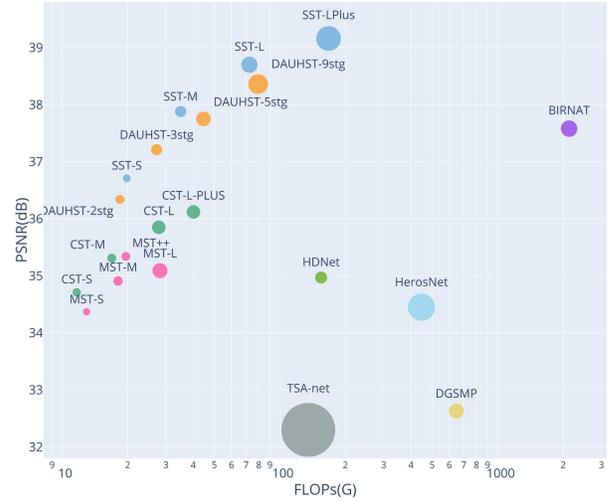}
   \caption{PSNR-Params-GFLOPs comparisons of our SST-ReversibleNet and SOTA HSI reconstruction methods.The vertical axis is PSNR (dB), the horizontal axis is GFLOPs (computational cost), and the circle radius is Params (memory cost).}
   \label{fig:1}
\end{figure}

\IEEEPARstart{W}{ith} rich and unique features \cite{cao2016computational}, hyperspectral images (HSIs) have been widely used for analysis and scene applications such as precision agriculture \cite{ishida2018novel}, national security \cite{udin2019uav}, environmental protection \cite{wright2019raman}, and astronomical observations \cite{de2015spectral}. In computer vision, HSIs can be extensively used for object tracking \cite{9684246, kim20123d}, material classification\cite{8960632}, feature extraction\cite{8935498}, and medical image analysis \cite{liu2019flexible}.

To obtain spectral images, traditional methods typically scan scenes along the 1D or 2D spatial dimension or along spectral channels, sacrificing time through multiple exposures to reconstruct the spectral data of the scene. Although traditional methods perform well in terms of spectral detection range and accuracy \cite{wang2021tensor}, they are unsuitable for dynamic detection and therefore consumer applications. Recently, researchers have exploited advances in compressed sensing (CS) theory to capture HSIs using Snapshot Compressed Imaging (SCI) systems \cite{du2009prism}, which compress information from snapshots along the spectral dimension into a single 2D measurement. Among current SCI systems, coded aperture snapshot spectral imaging (CASSI) \cite{wagadarikar2008single} stands out as a promising research direction.

Although the spectral data cube in the CASSI system is modulated by a coded mask and then dispersed, the complete data cube can be reconstructed from the redundant image information. Spectral reconstruction methods can be classified as traditional, Plug-and-Play (PnP), End-to-End (E2E) and Deep Unfolding Network (DU).

\IEEEpubidadjcol
Traditional methods perform reconstructions based on over-complete dictionaries or sparse spectral features that rely on hand-crafted priors and assumptions\cite{wang2016adaptive, zhang2019computational}. The main drawback of these traditional methods is the need to manually adjust parameters, resulting in poor robustness and slow reconstruction. In recent years, deep learning methods have demonstrated powerful capabilities in image generation and reconstruction \cite{arnab2021vivit}, such as image super-resolution, image denoising, and rain and fog removal \cite{10076399, liang2022details}, and have also been applied to spectral image reconstruction. PnP introduces a denoising module based on the traditional method, but with limited improvement in reconstruction speed and accuracy. The current SOTA methods all belong to E2E and DU. The E2E directly establishes the mapping between the measurement and truth data, and the DU uses a depth module to simulate the iterations in a convex optimization algorithm. Although both E2E and DU have achieved good performance, there are still limitations to the current methods.

1) The E2E method is similar to an open-loop control system where the measurements no longer guide the reconstruction process during the reconstruction, in addition to lacking interpretability and a DU-like iterative framework. As a result, E2E is inefficient in improving network performance by increasing network depth and limits the scope for improving accuracy.

2) The DU networks are based on convex optimisation algorithms, but require transposition and invertible operations on the operation matrix during iteration. These conditions limit the structure of the network module and impose requirements on the design of the coding mask, as described in the specific detailed work in Sec. II.

3) The denoising modules in the Transformer-based E2E methods and DU networks learn either the global self-similarity of the spectral dimension or the local correlation of the spatial dimension, ignoring the global correlation of the spectral cube in both the spectral and spatial dimensions.

The motivation of this paper is to find an interpretable E2E method with a structure similar to DU, but not subject to the constraints of convex optimization methods, thus bridging the gap between E2E and DU. Furthermore, we are looking for a mapping network that learns both the self-similarity of the transformer-based spectral dimension and the spatial global dependence of the transformer-based spatial dimension, taking into account memory consumption and computational complexity. 

To address the above issues, and inspired by the reversible nature of the optical path, we propose a framework based on the reversible optical path prior (Reversible-prior). Based on the learning of the residuals between the estimated and actual measurements of the reversible optical path, the new framework forms a closed loop that can effectively improve the reconstruction capability of the model, and the structure is shown in Fig. 2. Based on the new framework, a mapping network of Spectral-Spatial Transformer is designed to learn spectral and spatial self-similarity and global correlation using efficient spectral self-attention and spatial self-attention, respectively. We plug Spectral-Spatial Transformer into the reversible prior-based framework to establish a novel HSI reconstruction method, a Spectral-Spatial Transformer network based on reversible prior (SST-ReversibleNet). Finally, based on the unique design of the new framework, we propose a new reversible loss. Through the above proposed and improved methods, we establish a series of effective small-to-large SST-ReversibleNet families (SSTs), which outperform the state-of-the-art (SOTA) methods by a very large margin, as shown in Fig. 1.

Our contributions can be summarized as follows:

1) We propose a new framework that bridges the gap between E2E and DU, allowing E2E methods to have the interpretability and iterative capabilities of DU. In addition, we design a new reversible loss based on the new framework.

2) We present a Spectral-Spatial transformer module that can balance the parameters and reconstruction accuracy without deepening the depth of the module.

3) Our SST-ReversibleNet dramatically outperforms SOTA methods by a large margin while requiring cheaper computational and memory costs. Besides, SST-ReversibleNet yields more visually satisfying results in real-world HSI reconstruction.

\section{Related Works}
\label{sec:related work}
\subsection{Methods of HSI reconstruction}

\textbf{End-to-end method} The E2E method has a powerful mapping capability by directly finding strong mapping relationships between measurements and spectral cubes, so the network structure is concise and diverse. E2E can be divided into Convolutional Neural Networks (CNN-based) and transformer-based networks. CNN-based networks \cite{wang2015dual, yang2021associating, meng2020end, miao2019net}, such as TSA-net \cite{meng2020end}, learn local spatial correlations to reconstruct data, which has the advantage of fast inference, but tends to lose global features. Similarly, transformer-based networks use spectral self-attention to learn global similarity in spectral dimensions, or combine CNNs in space to compensate for local spatial information. However, the E2E networks all ignore how CASSI systems work and lack theoretical interpretability and flexibility.

\textbf{Deep unfolding network} The DU uses multi-stage network iterations to map measurements down a gradient into the HSI cube. DUs are derived from convex optimization algorithms, Half Quadratic Splitting (HQS), Alternating Direction Method of Multipliers (ADMM) and Proximal Gradident Descent (PGD) are common optimization algorithms with strong interpretability. These methods typically decompose the objective function into a data fidelity term and a regularized decoupling term, producing iterative schemes consisting of alternating solutions to a data subproblem and a prior subproblem. However, the optimization-based approach has some conditional constraints in the solution process, and as in the HQS expansion framework, the 2-stage iterative process can be described as \cite{cai2022degradation}:

\begin{equation}
    x_{k+1}=\left ( \Phi ^{\mathrm {T}  }\Phi  + \mu \mathrm {I}\right )^{-1}\left ( \Phi ^{\mathrm {T}}y +\mu z_{k}   \right ) 
  \label{eq:HQS1}
\end{equation}

\begin{equation}
    z_{k+1} = arg \underset{z}{min}\frac{1}{2\left ( \sqrt{\tau _{k+1}/\mu _{k+1}  }  \right )^{2}  }\left \| z-x_{k+1}  \right \|^{2} +R\left ( z \right )
  \label{eq:HQS2}
\end{equation}

where $\mathrm {I}$ is an identity matrix. $\Phi$ is a fat matrix. $X_{k+1} $ and $z_{k+1} $ are two subproblems of (k+1)-stage. $\mu$, $\mu _{k+1}$ are hyperparameters. $R\left (.\right )$ is a mapping function. It is clear from the formula that the optimization formula is valid on the premise that $\left ( \Phi ^{\mathrm {T}  }\Phi  + \mu \mathrm {I}\right )^{-1}$ is invertible. In addition, operations such as transpose multiplication $\Phi ^{\mathrm {T}  }\Phi$ and $\Phi ^{\mathrm {T}}y$ are involved in the operation.

\subsection{3D cube feature extraction module}

Both E2E and DU require feature extraction in the measurement space. Much of the previous work has revolved around extracting local spatial information using CNN \cite{liu2021swin, zhu2021deformable}, but these CNN-based models have limitations in capturing long-range spatial dependencies and modelling non-local self-similarity. Recently, the emerging Transformer has provided a solution to address the shortcomings of CNN. MST \cite{cai2022mask} proposed the first spectral transformer-based model for HSI reconstruction. MST treats spectral maps as tokens and computes self-attention along the spectral dimension. In addition, TSA-Net \cite{meng2020end} uses transformer-based spectral modules and CNN-based spatial modules to learn a non-linear mapping from the 2D measurement to the 3D hyperspectral cube. All these approaches ignore the global correlation in the spatial dimension.

\section{Model of CASSI System}
\label{sec:Model of CASSI System}

\begin{figure}[t]
  \centering
  \includegraphics[width=1\linewidth]{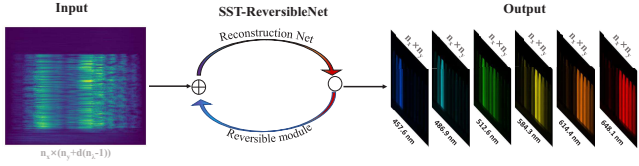}
  \caption{Schematic diagram of reversible optical path. According to the principle of reversible optical path, our network also includes two stages: forward and reverse.}
  \label{fig:2}
\end{figure}

\begin{figure}[!t]
  \centering
  \includegraphics[width=1\linewidth]{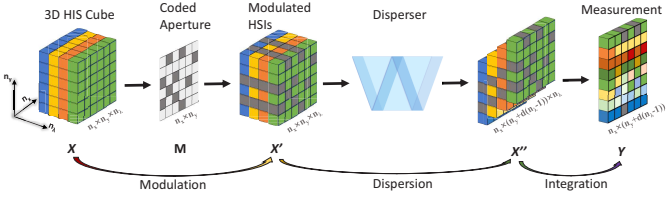}
  \caption{Imaging process of CASSI.}
  \label{fig:3}
\end{figure}

In CASSI system, the 3D hyperspectral cube is modulated via a coded mask and then dispersed by a dispersive prism (Fig. 3). Mathematically, considering a 3D HSIs cube, denoted by $X\in \mathbb{R} ^{n_{x}\times n_{y}\times c }$, where $n_{x}$, $n_{y}$, $c$ represent the HSIs’s height, width, and number of wavelengths. $M\in \mathbb{R} ^{n_{x}\times n_{y} }$ denoted a pre-defined mask. For each wavelength $m=1,2\cdots c$, the spectral image is modulated, and we can express it as:

\begin{equation}
 X^{\prime } \left (:,:,m\right ) = X\left (:,:,m\right ) \odot M
  \label{eq:CASSI0}
\end{equation}
Where $X^{\prime }\in \mathbb{R} ^{n_{x}\times n_{y}\times c}$ denotes the modulated spectral data-cube, and $\odot$ denotes the element-wise multiplication.
After passing the dispersive prism, $X^{\prime }$ becomes tilted and is considered to be sheared along the y-axis. We use $X^{\prime \prime } \in \mathbb{R} ^{ n_{x}\times \left ( n_{y}+  d\left ( c-1  \right ) \times c  \right ) } $ to denote the dispersed HSIs cube, where d represents the shifting step. We assume $\lambda _{c}$ to be the reference wavelength, i.e., $X^{\prime \prime } \left ( :,:,m \right )$ is not sheared along the y-axis. Then we have
\begin{equation}
X^{\prime \prime } \left ( x, y,m \right ) =X^{\prime } \left ( x,y+ d_{m},m\right )
  \label{eq:CASSI1}
\end{equation}
where (x, y) represents coordinates of a point on the 3D HSI, $d_{m}$ represents the spatial shifting of the m-th channel on $^{\prime \prime }$. Finally, the captured 2D compressed measurement $Y\in \mathbb{R} ^{n_{x}\times \left ( n_{y}+d\left ( c -1 \right )   \right )  }$ can be obtained by:

\begin{equation}
Y=\sum_{m=1}^{c } X^{\prime \prime } \left ( :,:,m \right ) +G
  \label{eq:CASSI2}
\end{equation}
where $G\in \mathbb{R} ^{n_{x}\times \left ( n_{y}+d\left ( c-1  \right )   \right )  }$ is the random noise generated by the photon sensing detector during the measurement.

\begin{figure*}
  \centering
   \includegraphics[width=0.8\linewidth]{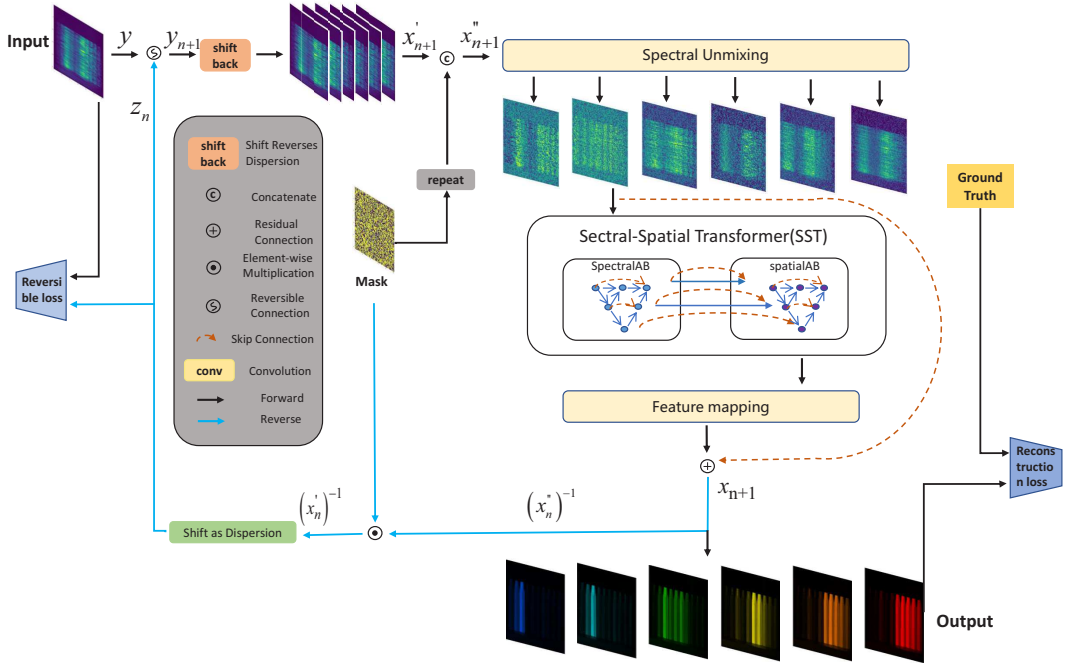}
   \caption{Diagram of the framework structure based on the reversible prior, the upper half is the forward process, reconstruction subnet including unmixing block, spectral-spatial transformer and mapping block. The blue line in the lower half indicates the inverse process, the reversible module. The reconstruction subnet corresponds to the inverse of the CASSI optical path and the reversible module corresponds to the forward direction of the CASSI optical path, both directions allowing the SST to form a closed-loop iterative capability. A reversible loss is proposed on the inverse of the network.}
   \label{fig:4}
\end{figure*}

\section{Proposed Method}

\subsection{Overall architecture based on reversible prior}

Previous E2E methods look for violent mapping relations to obtain the solution to Eq. (5) in a single pass, which means that they only have the upper half of the process (Reconstruction Net) of  Fig. 2. The single irreversible process also means that end-to-end methods cannot fine-tune the inference results, leading to a partial reduction in the performance of the model. According to the principle of optical path reversibility, it is easy to project the 3D cube of the reconstruction results back into the 2D measurement space relative to its inverse process. The construction of the residuals of the measured and reprojected data and the fine-tuning of the gap between the last learned data and the true value based on the residuals is the main difference between our network and the E2E and the DU.  The overall architecture of SST-ReversibleNet is shown in  Fig. 4. and is divided into a reversible module and a reconstruction subnet, which are represented as follows.

\begin{equation}
z_{n} = \mathcal{G}\left ( x_{n}  \right ) 
  \label{eq:residuals}
\end{equation}

\begin{equation}
 x_{n+1}=\mathcal{F}_{n+1} \left (y- z_{n} \right ) +  x_{n}
  \label{eq:Mapping}
\end{equation}
where $y$ is the actual measurement from the CCD camera, $\mathcal{G}$ is the mapping of the spectral 3D cube to the 2D measurement, $\mathcal{F}$ is the mapping from the input to the spectral 3D cube, $z_{n}$ is the output of the $n$-stage inverse process, and $x_{n+1}$ is the reconstruction result of the $(n+1)$-stage.

According to the number of stages $n$ of the output $x_{n}$, we establish four SST-ReversibleNet with small, medium, large and extremely large parameter sizes and computational costs: SST-S (n=1), SST-M(n=2), SST-L (n=4), SST-Lplus (n=9). In SST-S, we use reversible prior between SpatialAB and SpectialAB, while in other networks, we only use reversible prior between two SST modules.

\subsection{Reversible module}

The implementation of the inverse process is based on the output of the spectral reconstruction network at the n-th stage to obtain the predicted value of the spectral cube $x_{n}$. According to Eq. (3) and Eq. (4), the predicted value $\left ( {x_{n}}''\right )^{-1} = x_{n} $ can be projected back into the measurement space after mask encoding, dispersion and blending. As shown in the blue line in  Fig. 4, $\left ( {x_{n}}''\right )^{-1}$, $\left ( {x_{n}}' \right )^{-1} $ correspond to the inverse predicted values of ${x_{n}}''$ and ${x_{n}}'$ in the forward process, respectively, and the inverse process is described as:

\begin{equation}
 \left ( {x_{n}}' \right )^{-1}\left (x,y,:\right )=\left ( {x_{n}}'' \right )^{-1}\left (x,y,:\right ) \odot M
  \label{eq:ReversibleNet2}
\end{equation}

\begin{equation}
 z_{n}\left ( x,y\right )=\sum_{m=1}^{c }\left ( {x_{n}}' \right )^{-1}\left ( x,y+d_{m},m \right )
  \label{eq:ReversibleNet3}
\end{equation}

After obtaining $z_{n}$, our reconstruction network reconstructs a residual $y-z_{n}$ of the measurements $y$ and feeds it again into the reconstruction network to relearn the mapping of $y-z_{n}$ to $x_{Truth}-x_{n}$. The input $y_{n+1}$ to the reconstruction subnet is represented as

\begin{equation}
y_{n+1}=\begin{cases}
 y  & \text{ if } n=0 \\
 y-z_{n}    & \text{ otherwise}
\end{cases}
  \label{eq:predicted}
\end{equation}

\begin{figure*}
  \centering
   \includegraphics[width=0.8\linewidth]{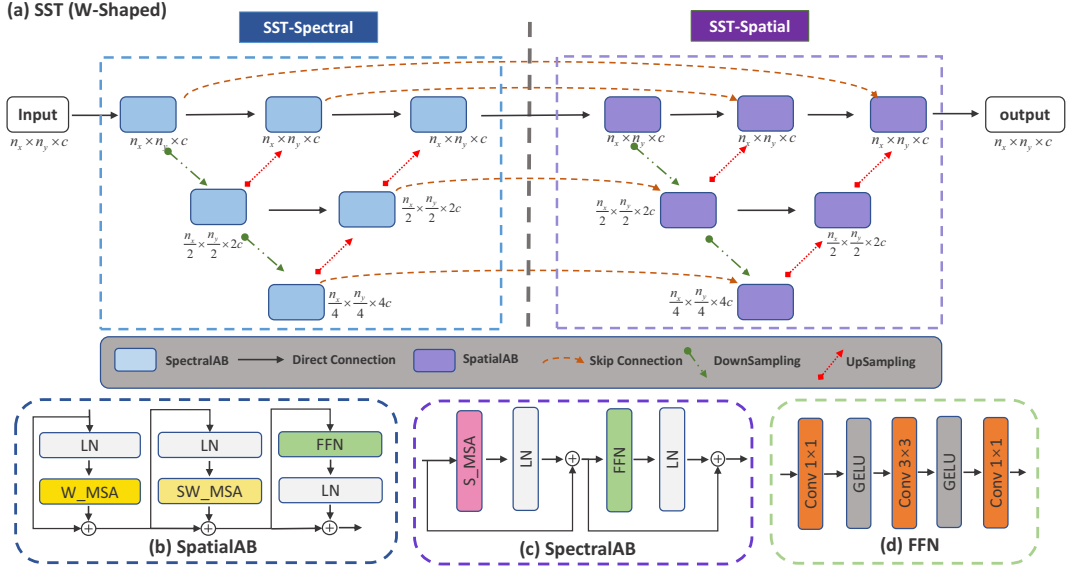}
   \caption{Diagram of Spectral-Spatial Transformer. (a) SST adopts a W-shaped structure. (b) SpatialAB consists of a Window Multi-head-Self-Attetion (MSA), a Shifted-Window-MSA, an FFN, and three layer normalization. (c) SpectralAB consists of a Spectral-MSA, an FFN, and two layer normalization.(d) Components of FFN. }
   \label{fig:5}
\end{figure*}

\subsection{Reconstruction subnet}
The function of the Reconstruction subnet is to establish a mapping between the different inputs and outputs, mainly consisting of a spectral-spatial transformer. In addition, considering that the measurement space is a compressed 2D space and that there is aliasing of data from different channels, we introduced a module for unaliasing and feature mapping before the mapping network input and after the output.

Given the measurement after initialization $x_{n+1}\in \mathbb{R} ^{ n_{x}\times \left ( n_{y}+  d_{m} \right )}$  , firstly, we divide the aliased data into input signals with different wavebands according to the backlight propagation. Obtained the initialized signal $X_{n+1}^{\prime}\in \mathbb{R }^{n_{x}\times n_{y}\times c} $ as:

\begin{equation}
x_{n+1}^{\prime}\left ( x,y ,m\right ) =y_{n+1} \left ( x,d_{m}:d_{m}+h\right),m=1,2,\dots ,c
  \label{eq:ReconstructionSubnet}
\end{equation}
where $x$, $y$ are the spatial coordinates of a point on the 3D cube, $d_{m}$ is the offset of the spectral image on the m-channel, and $h$ is the height of the 3D cube.

\noindent\textbf{Spectral Unmixing.} Subsequently, we use the prior of the mask to guide the input to unmix by passing the shifted y concatenated with mask M, then through convolution with $conv 1\ast1$ kernel to back to input signal $ X_{n}^{\prime\prime} \in \mathbb{R} ^{n_{x}\times n_{y}\times 2c} \longrightarrow {X}_{n} \in \mathbb{R }^{n_{x}\times n_{y}\times c} $ . The spectral unmixing is realized through the convolution layer with varying sizes to solve the aliasing problem under different receptive fields($conv 3\ast3$, $conv 5\ast5$, $conv 7\ast7$).

\noindent\textbf{Spectral-Spatial Transformer.} The proposed reconstruction subnet aims to reconstruct high-quality HSIs from the spectral images after unmixing. We use a W-shaped spectral-spatial transformer module (SST, Fig. 5), which is composed of encoding and decoding of spectral features and the encoding and decoding between spatial channels.

Both the SST-Spectral and SST-Spatial use an encoder-decoder unet-like architecture, which are connected by a series of nested dense SpatialAB and SpectralAB blocks, respectively. This architecture is designed to fuse the gaps between the feature maps of the encoder and decoder for the same feature in different dimensions.

\noindent\textbf{Spectral-Spatial-Wise Multi-Head Self-Attention.} The Cube of the spectrum has a spatial correlation in the spatial dimension, which is related to the  target's properties and the surface's reflectivity. While, in the spectral dimension, the continuity of the spectrum determines that the adjacent spectra are similar, and the farther the spectral distance is, the more ranges are complementary. And since $W=H>>M$, capturing spectral-wise interactions will be less cost-effective than modeling spatial-wise correlations. However, when the model reaches a certain scale, a single method cannot continue to mine the information of spectral Cube.

Our SpatialAB and SpectralAB are inspired by Swin-Transformer and MST respectively. SpectralAB is consistent with MSAB in MST\cite{cai2022mask}. SpectralAB's goal is to treats each spectral feature map as a token and calculates self-attention along the spectral dimension. The input $x_{n}\in \mathbb{R} ^{n_{x} \times n_{y} \times c} $ is reshaped into tokens $x\in \mathbb{R} ^{n_{x}n_{y}\times c}$. Than $x$ is  linearly projected into $query Q$, $key K$, $value V \in \mathbb{R}^{n_{x}n_{y}\times c} $, and $Q=xW^{Q} ,K=xW^{K} ,V=xW^{V} $, where $W^{Q} ,W^{K} ,W^{V} \in \mathbb{R}^{c\times c}$. Subsequently, $Q$, $K$, and $V$ into N heads along the spectral channel dimension: $Q=\left[Q_{1},\dots ,Q_{N}\right]$, $k=\left[k_{1},\dots,k_{N}   \right ]$, $v=\left[ v_{1},\dots ,v_{N}\right]$. Therefore, the formula for each $head_{j}^{Spectral}$ and SpectralAB is:

\begin{equation}
head_{j}^{Spectral} =Softmax\left ( \sigma _{j} Q_{j}K_{j}^{T} \right )V_{j}   
  \label{eq:head}
\end{equation}
\begin{equation}
SpectralAB\left ( X \right ) =\underset{j=1}{\overset{N}{concat }}  \left (head_{j} \right )W+f\left ( V  \right ) 
  \label{eq:SpectralAB}
\end{equation}
where $K_{j}^{T}$ denotes the transposed matrix of $K_{j}$. $W\in \mathbb{R} ^{c\times c} $are learnable parameters, $f\left ( \cdot  \right ) $ is the function to generate position embedding.

 SpatialAB makes improvements  based on Swin-Transformer\cite{liu2021swin}. We remove Avgpooling blocks, and add the Feature Forward Network (FFN) module and a LayerNorm layer (Fig. 5). SpatialAB's goal is to treats each local spatial feature map as a token and calculates self-attention along the spatial dimension. The input $x_{n}\in \mathbb{R} ^{n_{x} \times n_{y} \times c} $ is reshaped into tokens  $x\in \mathbb{R} ^{\frac{n_{x} }{s}\cdot \frac{n_{y}}{s} \times s\cdot s\times c}$, s represents the window-size (set to 8 by default) of each window. Than $x$ is  linearly projected into query $Q$, key $K$, and value $V$, and $W^{Q} ,W^{K} ,W^{V} \in \mathbb{R}^{s\times s}$. Then, the next module adopts a windowing configuration that is shifted from that of the preceding layer, by displacing the windows by $\left ( \left \lfloor \frac{s}{2}  \right \rfloor ,\left \lfloor \frac{s}{2}  \right \rfloor \right ) $ pixels from the regularly partitioned windows.The formula for spatial Attention is as follows:
\begin{equation}
Attention=SoftMax(QK^{T}/\sqrt{d}+B)V
  \label{eq:SpatialAB}
\end{equation}
Where $Q,K,V \in \mathbb{R}^{s^{2} \times c }$ are the query, key and value matrices; $B$ is the relative position embedding, $B\in \mathbb{R}^{ s^{2} \times s^{2} }$; $d$ is the $query/key$ dimension, and $s^{2}$ is the number of patches in a window.

\begin{figure*}
  \centering
   \includegraphics[width=0.95\linewidth]{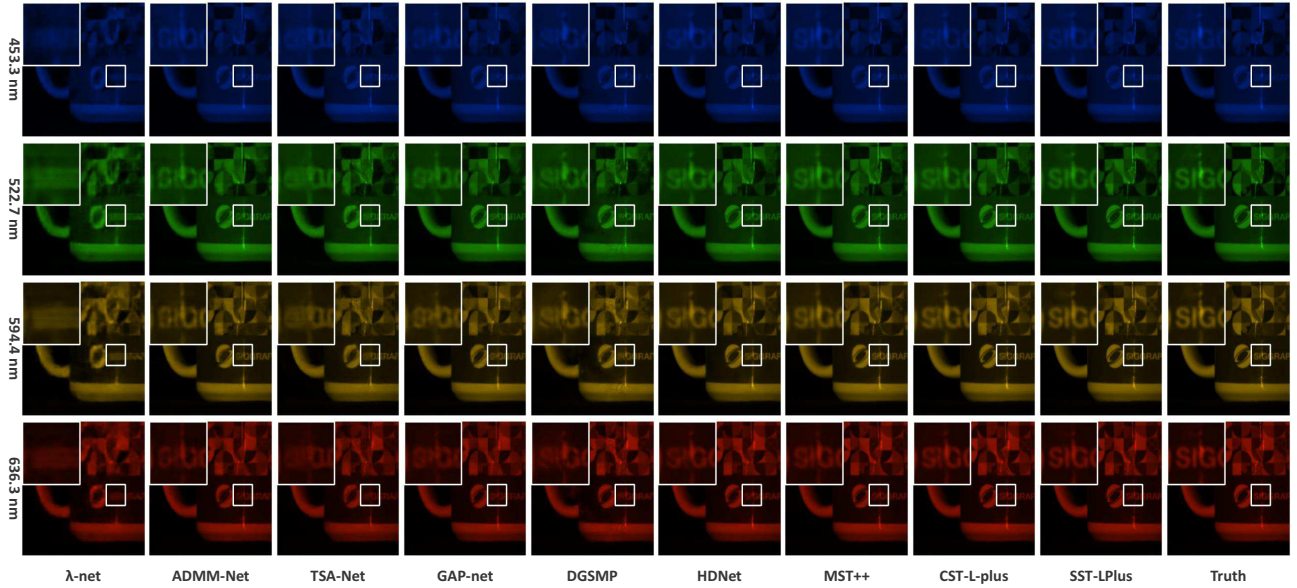}
   \caption{Visual comparisons of our SST-ReversibleNet and other SOTA methods of Scene 5 with 4 out of 28 spectral channels on the KAIST dataset.}
   \label{fig:6}
\end{figure*}

\begin{figure}
  \centering
   \includegraphics[width=1\linewidth]{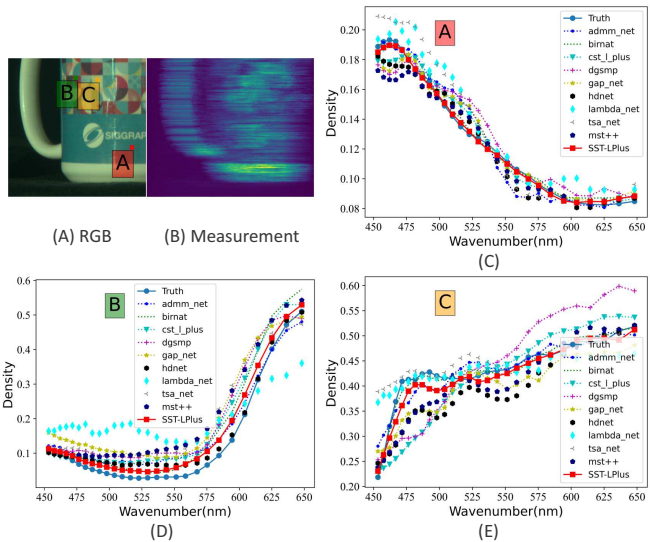}
   \caption{Spectral curves of the SOTA methods in Fig. 6 on the randomly selected regions A, B and C.}
   \label{fig:7}
\end{figure}

\subsection{Loss Function.}

Our network has reversible module and reconstruction subnet, so our loss includes outputting and reversible loss. The outputting loss is calculated as the L2 loss of $x_{out}-x_{truth}$. The reversible loss calculation $x_{out}$ is mapped back to the CCD under the nature of the reversible optical path to obtain the L2 loss of the $ \mathcal{G}\left ( x_{out}  \right ) $ value to the actual measurement $y$. We defined the loss function as follows:
\begin{equation}
\mathcal{L} =\left \| x_{out}- x_{truth}  \right \|_{2}^{2}  + \xi \cdot \left \| \mathcal{G}\left ( x_{out} \right ) -y \right \|_{2}^{2} 
  \label{eq:lossout}
\end{equation}
where $x_{out}$ is the final predicted values of the network, $\mathcal{G}$ represents the process of mask coding and dispersion of predicted values, $y$ is the measurement of CCD. $\xi$ is the penalty coefficient, which is set to 0.2 by default.

\section{Experiments}
\subsection{Experiment Setup}

In our implementation, the number of spectral channels c is 28, wavelengths from 450 nm to 650 nm. We perform experiments on both simulation and real HSIs datasets.

\noindent\textbf{Simulation HSIs Data.} We use two simulation hyperspectral image datasets, CAVE\cite{park2007multispectral} and KAIST\cite{choi2017high}. CAVE dataset is composed of 32 hyperspectral images at a spatial size of 512 $\times$ 512. KAIST dataset consists of 30 hyperspectral images at a spatial size of 2704 $\times$ 3376. Following the schedule of TSA-Net, we adopt CAVE as the training set. 10 scenes from KAIST are selected for testing.

\noindent\textbf{Real HSIs Data.} We use the real HSIs dataset collected by the CASSI system developed in TSA-Net\cite{meng2020end}.

\noindent\textbf{Evaluation Metrics.} We adopt peak signal-to-noise ratio (PSNR) and structural similarity (SSIM)\cite{wang2004image} as the metrics to evaluate the HSI reconstruction performance.

\noindent\textbf{Implementation Details.} We implement SST-ReversibleNet in Pytorch. Our SST-S, SST-M, SST-L are trained on 1 $\times$ RTX 3090 GPU, and SST-LPlus is trained on 2 $\times$ RTX 3090 GPUs. We adopt Adam optimizer ($\beta_{1} = 0.9$ and $\beta_{2} = 0.999$) for 300 epochs. The learning rate is set to $4\times 10^{-4}$ in the beginning and is halved every 50 epochs during the training procedure. The metrics of PSNR and SSIM are employed to evaluate the reconstruction quality.

\begin{table*}[]
\begin{center}
\caption{Comparisons between SSTs and SOTA methods on 10 simulation scenes (S1$\sim$S10). Params, GFLOPS, PSNR and SSIM are reported.}
\begin{tabular}{ccccccccccccccc}
\hline
\multicolumn{2}{l}{scene}         & S1     & S2     & S3     & S4     & S5     & S6     & S7     & S8     & S9     & S10    & Avg & Params & GFLOPs  \\
\hline
$\lambda$-net \cite{miao2019net}   & PSNR & 32.50  & 31.23 & 33.89 & 40.28 & 29.86 & 30.27 & 30.33 & 28.98 & 31.98 & 28.36 & 31.77 & 62.6 & 118.0 \\
                           & SSIM & 0.892 & 0.854 & 0.930  & 0.965 & 0.889 & 0.893 & 0.975 & 0.880  & 0.891 & 0.834 & 0.890  \\
\hline
ADMM-Net \cite{ma2019deep} & PSNR & 34.12  & 33.62 & 35.04 & 41.15 & 31.82 & 32.54 & 32.42 & 30.74 & 33.75 & 30.68 & 33.58 & 4.27 & 78.58 \\
                           & SSIM & 0.918 & 0.902 & 0.931  & 0.966 & 0.922 & 0.924 & 0.896 & 0.907  & 0.915 & 0.895 & 0.918  \\
\hline
TSA-Net \cite{meng2020end}  & PSNR & 32.95 & 31.69 & 33.01 & 41.24 & 30.12 & 31.89 & 30.75 & 29.89 & 31.61 & 29.9  & 32.30 & 44.2 & 135.2 \\
                           & SSIM & 0.913 & 0.884 & 0.932 & 0.975 & 0.911 & 0.929 & 0.895 & 0.912 & 0.920  & 0.890  & 0.916 \\
\hline
GAP-net \cite{meng2020gap}  & PSNR & 26.82 & 22.89 & 26.31 & 30.65 & 23.64 & 21.85 & 23.76 & 21.98 & 22.63 & 23.10  & 24.36 & 4.27 & 84.47 \\
                           & SSIM & 0.754 & 0.610  & 0.802 & 0.852 & 0.703 & 0.663 & 0.688 & 0.655 & 0.682 & 0.584 & 0.669 \\
\hline
DGSMP \cite{huang2021deep}    & PSNR & 33.26 & 32.09 & 33.06 & 40.54 & 28.86 & 33.08 & 30.74 & 31.55 & 31.66 & 31.44 & 32.63 & 3.76 & 646.7 \\
                           & SSIM & 0.915 & 0.898 & 0.925 & 0.964 & 0.882 & 0.937 & 0.886 & 0.923 & 0.911 & 0.925 & 0.917 \\
\hline
DIP-HSI \cite{meng2021self}    & PSNR & 32.68 & 27.26 & 31.30 & 40.54 & 29.79 & 30.39 & 28.18 & 29.44 & 34.51 & 28.51 & 31.26 & 33.9 & 64.42 \\
                           & SSIM & 0.890 & 0.833 & 0.914 & 0.962 & 0.900 & 0.877 & 0.913 & 0.874 & 0.927 & 0.851 & 0.894 \\
\hline
BIRNAT \cite{cheng2022recurrent}   & PSNR & 36.79 & 37.89 & 40.61 & 46.94  & 35.42 & 35.30 & 36.58 & 33.96 & 39.47 & 32.80 & 37.58 & 4.35 & 2131 \\
                           & SSIM & 0.951 & 0.957  & 0.971 & 0.985 & 0.964 & 0.959 & 0.955 & 0.956 & 0.970 & 0.938 & 0.960 \\
\hline
HDNet \cite{hu2022hdnet}     & PSNR & 35.14 & 35.67 & 36.03 & 42.30  & 32.69 & 34.46 & 33.67 & 32.48 & 34.89 & 32.38 & 34.97 & 2.37 & 154.8 \\
                           & SSIM & 0.935 & 0.940  & 0.943 & 0.969 & 0.946 & 0.952 & 0.926 & 0.941 & 0.942 & 0.937 & 0.943 \\
\hline
MST-L \cite{cai2022mask}    & PSNR & 35.29 & 35.48 & 36.72 & 42.68 & 32.55 & 34.67 & 33.53 & 32.50  & 34.98 & 32.45 & 35.09 & 3.66 & 28.15\\
                           & SSIM & 0.945 & 0.944 & 0.956 & 0.980 & 0.947 & 0.957 & 0.929 & 0.953 & 0.948 & 0.945 & 0.950 \\
\hline
MST++ \cite{cai2022mst++}     & PSNR & 35.53 & 35.68 & 35.99 & 42.78 & 32.71 & 35.14 & 34.24 & 33.30  & 35.13 & 32.86 & 35.34 & 1.33 & 19.64\\
                           & SSIM & 0.946 & 0.946 & 0.954 & 0.977 & 0.949 & 0.959 & 0.938 & 0.957 & 0.951 & 0.948 & 0.953 \\
\hline
CST-L \cite{lin2022coarse}    & PSNR & 35.96 & 36.84 & 38.16 & 42.44 & 33.25 & 35.72 & 34.86 & 34.34 & 36.51 & 33.09 & 36.12 & 3.00 & 40.10\\
                           & SSIM & 0.949 & 0.955 & 0.962 & 0.975 & 0.955 & 0.963 & 0.944 & 0.961 & 0.957 & 0.945 & 0.957 \\
\hline
DAUHST-2st \cite{cai2022degradation} & PSNR & 35.93 & 36.70 & 37.96 & 44.38 & 34.13 & 35.43  & 34.78 & 33.65 & 37.42  & 33.07 & 36.34 & 1.40 & 18.44\\
                           & SSIM & 0.943 & 0.946 & 0.959 & 0.978 & 0.954 & 0.957  & 0.940 & 0.950 & 0.955  & 0.941 & 0.952 \\

DAUHST-3st \cite{cai2022degradation} & PSNR & 36.59 & 37.93 & 39.32 & 44.77 & 34.82  & 36.19 & 36.02 & 34.28  & 38.54 & 33.67 & 37.21 & 2.08 & 27.17\\
                           & SSIM & 0.949 & 0.958 & 0.964 & 0.980 & 0.961 & 0.963  & 0.950 & 0.956 & 0.963  & 0.947 & 0.959 \\

DAUHST-5st \cite{cai2022degradation} & PSNR & 36.92 & 38.52 & 40.51 & 45.09 & 35.33  & 36.56 & 36.28 & 34.74  & 38.71 & 34.27 & 37.75 & 3.44 & 44.61\\
                           & SSIM & 0.955 & 0.962 & 0.967 & 0.980 & 0.964 & 0.965  & 0.958 & 0.959 & 0.963  & 0.952 & 0.962 \\

DAUHST-9st \cite{cai2022degradation} & PSNR & 37.25 & 39.02 & 41.05 & 46.15 & 35.80  & 37.08 & \pmb{37.57} & 35.10  & 40.02 & 34.59 & 38.36 & 6.15 & 79.50\\
                           & SSIM & 0.958 & 0.967 & 0.971 & 0.983 & 0.969 & 0.970  & \pmb{0.963} & 0.966 & 0.970  & 0.956 & 0.967 \\
\hline
SST-S  & PSNR & 36.55 & 37.27 & 38.49 & 44.50 & 34.30 & 36.18 & 35.35 & 34.03 & 37.21 & 33.19 & 36.71 & 1.06 & 19.83\\
                           & SSIM & 0.953 & 0.955 & 0.964 & 0.984  & 0.960 & 0.966 & 0.948 & 0.959 & 0.955 & 0.949 & 0.959 \\

SST-M & PSNR & 37.32 & 38.60 & 40.76 & 45.73 & 35.56 & 37.01 & 36.44 & 34.69 & 38.63 & 34.09 & 37.88 & 2.11 & 35.03\\
                           & SSIM & 0.961 & 0.965 & 0.973 & 0.987 & 0.968 & 0.972 & 0.955 & 0.966 & 0.966 & 0.959 & 0.967 \\

SST-L & PSNR & 37.77 & 39.56  & 41.87 & 46.72 & 36.50 & 37.54 & 37.28 & 35.11 & 39.80 & 34.83 & 38.70& 4.25 & 72.52\\
                           & SSIM & 0.966  & 0.972 & 0.977 & 0.990 & 0.972 & 0.975 & 0.962 & 0.968 & 0.972 & 0.963 & 0.972 \\

SST-Lplus & PSNR & \pmb{38.24}  & \pmb{40.05} & \pmb{42.45} & \pmb{47.87} & \pmb{37.02} & \pmb{37.59} & 37.20 & \pmb{35.42} & \pmb{40.54} & \pmb{35.25} & \pmb{39.16}  & 9.64 & 167.5\\
                           & SSIM & \pmb{0.968} & \pmb{0.974} & \pmb{0.979} & \pmb{0.992} & \pmb{0.975} & \pmb{0.975} & 0.960 & \pmb{0.971} & \pmb{0.975} & \pmb{0.962} & \pmb{0.974} \\
\hline
\label{tab:1}
\end{tabular}
\end{center}
\end{table*}

\subsection{Simulation results}
We compare the Params, GFLOPs, PSNR, and SSIM of our SST-ReversibleNet with several SOTA HSI reconstruction algorithms, including $\lambda$-net\cite{miao2019net}, ADMM-Net\cite{ma2019deep}, TSA-Net\cite{meng2020end}, DIP-HSI\cite{meng2021self},  DGSMP\cite{huang2021deep}, BIRNAT\cite{cheng2022recurrent}, MST series\cite{cai2022mask, cai2022mst++}, CST series\cite{lin2022coarse}, HDNet\cite{hu2022hdnet}, and DAUHST series\cite{cai2022degradation} . The Params, GFLOPs are tested with the same settings (test size = 256 $\times$ 256), PSNR and SSIM results of different methods on 10 scenes in the simulation datasets are listed in Table I.

\begin{table}[]\scriptsize
\begin{center}
\caption{Ablation of use piror.}
\begin{tabular}{cccccc}
\hline
base-line & Use Piror & Params(M) & GFLOPs(G) & PSNR  & SSIM    \\
\hline
SST-S &  $\times$  & 1.03  & 17.98  & 33.52 & 92.2\% \\
SST-S & $\surd$ & 1.06   & 19.83  & 36.71 & 95.9\%\\
\hline
\label{tab:2}
\end{tabular}

\hspace*{\fill}
\caption{ Ablation of use reversible loss.}
\begin{tabular}{cccccc}
\hline
base-line & Use reprojection & Params & GFLOPs & PSNR  & SSIM    \\
 & loss & (M) & (G) &   &     \\
\hline
SST-S &  $\times$  & 1.06  & 19.83  & 36.66 & 95.7\% \\
SST-S & $\surd$ & 1.06   & 19.83  & 36.71 & 95.9\%\\
\hline
\label{tab:3}
\end{tabular}

\hspace*{\fill}
\caption{ Ablation of spectral-spatial Transformer structures.}
\begin{tabular}{cccccc}
\hline
base-line & shape of spectral & Params & GFLOPs & PSNR  & SSIM    \\
 & spatial Transformer & (M) & (G) &   & \\
\hline
SST-S &  Unet-like  & 1.01  & 16.99  & 35.42 & 95.3\% \\
SST-S & Unet++-like & 1.06   & 19.83  & 36.71 & 95.9\%\\
SST-M & W-shaped & 2.11   & 35.03  & 37.86 & 96.6\%\\
\hline
\label{tab:4}
\end{tabular}

\hspace*{\fill}
\caption{ Ablation of use spectralAB and spatialAB.}
\begin{tabular}{ccccccc}
\hline
base-line & Use & Use & Params & GFLOPs & PSNR  & SSIM    \\
 & spectralAB &spatialAB & (M) & (G) &   &     \\
\hline
A &  $\surd$ &  $\times$  & 0.97  & 19.61  & 34.83 & 93.2\% \\
B &  $\times$ & $\surd$ & 1.13   & 21.74  & 35.77 & 94.7\%\\
SST-S & $\surd$ & $\surd$ & 1.06   & 19.83  & 36.71 & 95.9\%\\
\hline
\label{tab:5}
\end{tabular}
\end{center}
\end{table}

\begin{figure*}
  \centering
   \includegraphics[width=0.95\linewidth]{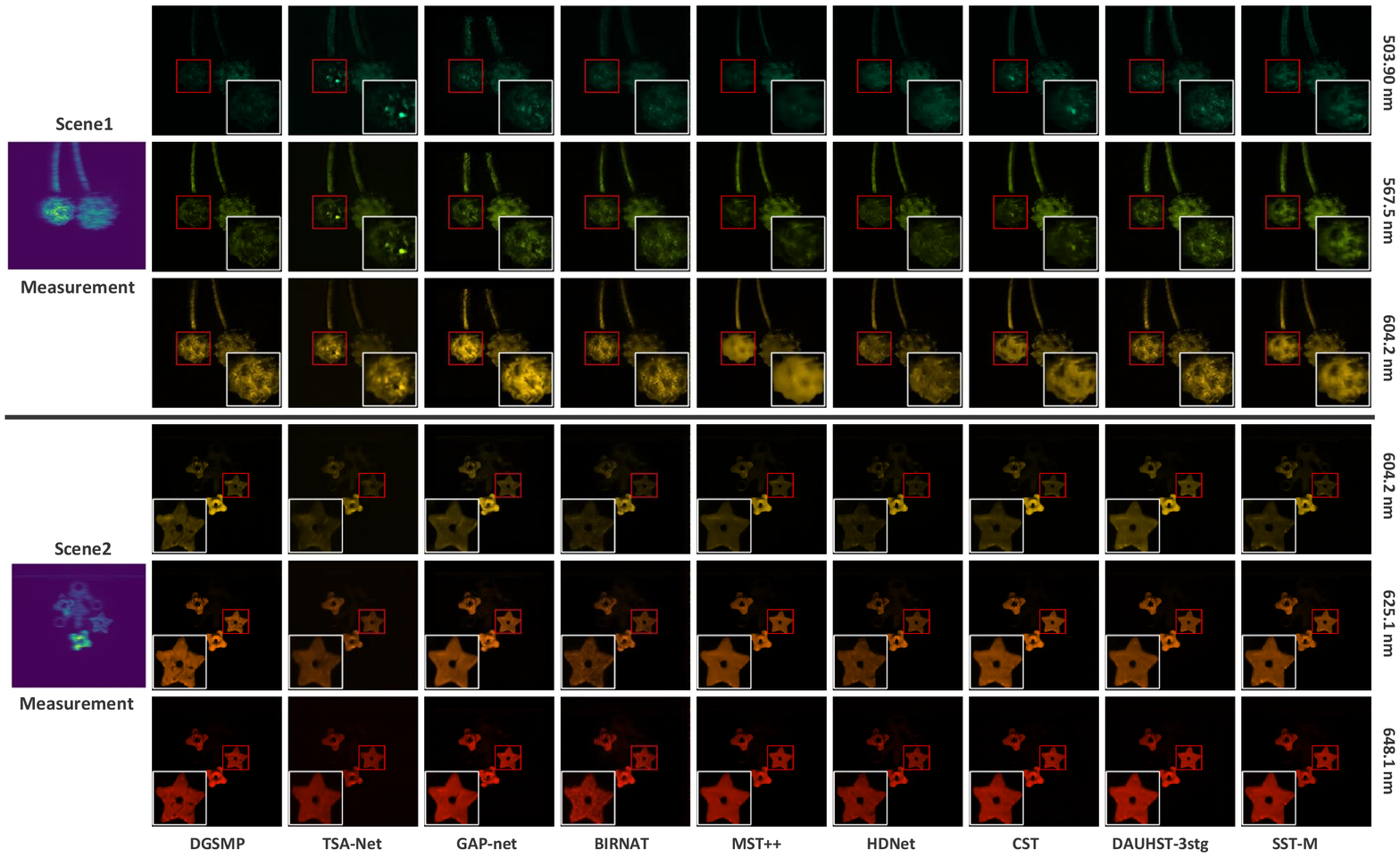}
   \caption{Real HSI reconstruction comparison of two Scenes. 6 out of 28 spectra are randomly selected.}
   \label{fig:8}
\end{figure*}

\textbf{(i)}Our best model SST-LPlus yields very impressive results, i.e., 39.16 dB in PSNR and 97.4\% in SSIM , which is more than 3 dB than the best PSNR of the SOTA published models, and the SSIM is more than 1.5\%. SST-LPlus significantly outperforms DAUHST-9stg, BIRNAT, MST++, MST-L, HDNet, TSA-net and $\lambda$-net of PSNR by 0.80, 1.58, 3.82, 4.07, 4.19, 6.86 and 7.39 dB, and 0.7\%, 1.4\%, 2.1\%, 2.4\%, 3.1\%, 5.8\% and 8.4\% improvement of SSIM, suggesting the effectiveness of our method.

Fig. 6 plots the visual comparisons of our SST-LPlus and other SOTA methods on Scene 5 with 4 (out of 28) spectral channels. The top-right part shows the zoomed-in patches of the white boxes in the entire HSIs, the reconstructed HSIs produced by SSTs have more spatial details and clearer texture in different spectral channels than other SOTA methods. In addition, as illustrated in Fig. 7, in A, B, and C three positions, although all the restoration algorithms can better describe the qualitative trend of spectral changes, the spectral curves of the SSTs have higher spectral accuracy and better perceptual quality.

\textbf{(ii)}It can be observed that our SST-ReversibleNet significantly surpass SOTA methods by a large margin while requiring much cheaper memory and computational costs. Compared with other Transformer-based method CST-L and MST-L, our SST-S outperforms CST-L\cite{lin2022coarse} by 0.59 dB but only costs 35.3\% (1.06/3.00) Params and 71.3\% (19.83/27.81) GFLOPs, and SST-S outperforms MST-L\cite{cai2022mask} by 1.62 dB, but only costs 29.0\% (1.06/3.66) Params and 70.4\% (19.83/28.15) GFLOPs. Likewise, our SST-M outperforms DAUHST-5stg\cite{cai2022degradation} by 0.13 dB but only costs 61.3\% (2.11/3.44) Params and 78.5\% (35.03/44.61) GFLOPs, and SST-M outperform CST-L-plu by 1.76 dB, but only costs 70.3\% (2.11/3.00) Params and 87.4\% (35.03/40.1) GFLOPs. More specifically, our SST-M acquire the equivalent SSIM (96.7\%) of DAUHST-9stg\cite{cai2022degradation} (the best model at present), but only costs 34.3\% (2.11/6.15) Params and 44.1\% (35.03/79.5) GFLOPs. In addition, our SST-L and SST-LPlus outperforms other competitors by very large margins. we provide PSNR-Params-GFLOPs comparisons of different reconstruction algorithms in Fig. 1.

\subsection{Real data results}
To verify the effect of the proposed method on the real data, five compressive measurements captured by the real spectral SCI system are utilized for testing. For fair comparisons, all of the methods are trained on the CAVE datasets using the fixed real mask with 11-bit shot noise injected. Fig. 8 plots the visual comparisons of the proposed SST-M and the existing SOTA method  DGSMP\cite{huang2021deep}, TSA-Net\cite{meng2020end}, GAP-net\cite{meng2020gap}, BIRNAT\cite{cheng2022recurrent}, MST++\cite{cai2022mst++}, HDNet\cite{hu2022hdnet}, CST\cite{lin2022coarse}, DAUHST\cite{cai2022degradation} . Our SST-S surpasses previous algorithms in terms of high-frequency structural detail reconstruction and real noise suppression. In Scene 2, the proposed method is able to restore more texture and detail, especially at the edges of flowers.

\subsection{Ablation study}
To evaluate the contribution of different components in the proposed SST-ReversibleNet, ablation study is conducted on the CAVE and KAIST datasets. We mainly focus on the four components, i.e. whether to use the reversible prior, whether to use the reversible loss, the structural shape of the feature extraction network, and the effect of the combination of spectral self-attentive blocks (spectralAB) and spatial self-attentive blocks (spatialAB) on the model. Table II to Table V show the results of the comparison between PSNR and SSIM at different settings. In Table V, we build two networks A and B with similar number of parameters and GFLOPs as SST-S. In this case, A uses only the self-attention of the spectral channels and B only looks for correlations in the spatial dimension.

The results show that 1) the impact of reversible prior on the model is crucial, comparing two networks with similar number of parameters for deepening the network depth and using reversible prior in Table II, the use of reversible prior can effectively improve the reconstruction ability of the model, with PSNR and SSIM improving by 3.19 dB and 3.7\%, respectively. 2) The reversible loss can be constraint on the model, and without changing the number of parameters or operations, the PSNR and SSIM are able to improve by 0.05db and 0.2\% respectively. 3) The shape of the network structure also has a more obvious improvement on the reconstruction effect, the W-shaped structure can improve the PSNR and SSIM by 2.44 db/1.3\% and 1.15 db/0.7\%, respectively, compared to Unet-like and Unet++like shape. 4) The spectral-space transformer is a huge advantage over the spectral-transformer and spatial-transformer, especially the results for model B vs. SST-s. Model B has 7\% higher number of parameters and 10\% higher GFLOPs than SST-S, but the reconstruction results are 0.94 db less than SST-S.

\begin{figure*}
  \centering
   \includegraphics[width=0.85\linewidth]{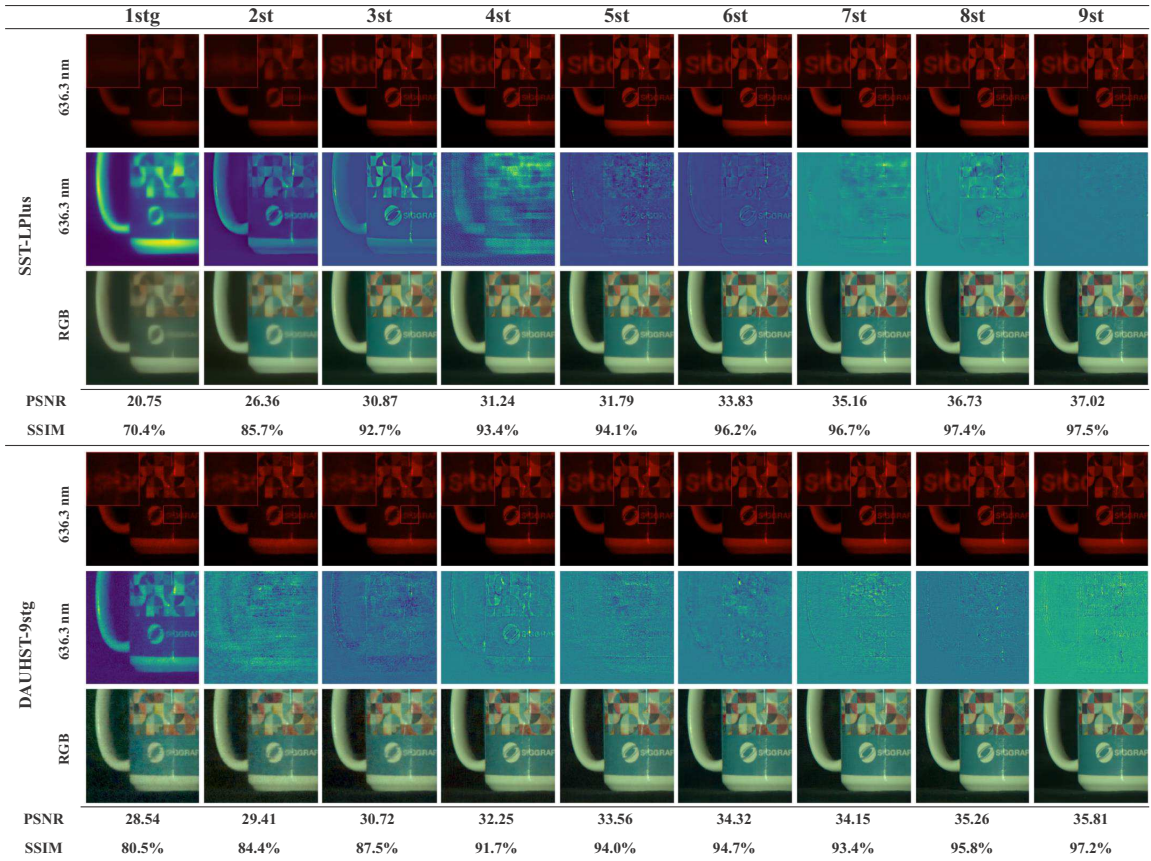}
   \caption{Comparison graph of E2E iterative method (SST-LPlus) and DU (DAUHST-9stg) in different stages. Comparison of the differences on the 636.3 nm spectral image, the 636.3 nm spectrally learned feature map, the RGB image, the PSNR and the SSIM from 1stg to 9stg.}
   \label{fig:9}
\end{figure*}

\subsection{Difference with DU}
To explore the differences between our iterative method and the DU, we compared the iterative process of SST-LPlus and DAUHST-9st on the simulated dataset. A randomly selected scene is visualised in both spectral channels and RGB changes. In addition, below the 636.3 nm visualization image, we have extracted the changes at different stages of each feature learning. Below the image we list the PSNR and SSIM changes from 1stg to 9stg, as shown in Fig. 9.

Visualizing the analysis of the results, we believe that the reversible framework benefits from the learning of residuals to effectively improve the learning of features, with DAUHST learning a large number of global features at 1stg and fine-tuning from 2stg onwards. Although SST only starts fine-tuning at 4stg, our SST learns more global features from 1stg to 4stg in the early stage and achieves results beyond DAUHST at 7stg to 9stg. We therefore believe that the feature learning capabilities of SST and DU are not the same.

\section{Conclusion}
Inspired by the reversible light path, this paper proposes a novel SST-ReversibleNet for CASSI. The new framework significantly improves the reconstruction metrics and can be used for other algorithms. We use a W-shaped spectral-spatial transformer module to improve spatial and spectral feature extraction. In addition, we design a reversible loss. With these novel techniques, we establish a set of highly efficient SST-ReversibleNet models. Quantitative experiments show that our method outperforms SOTA algorithms by a wide margin, even when using significantly cheaper parameters and GFLOPs. However, as presented in Fig. 8, our framework is not as effective on real datasets as it is on simulated datasets, and we believe that there is a lack of noise estimation in the reversible module, but there is currently a lack of relevant datasets. Therefore, our future work will be to construct noisy datasets based on real scenarios and to optimise our framework.


\bibliographystyle{IEEEtran}
\bibliography{IEEEabrv,mybibfile.bib}

\vfill

\end{document}